\documentclass{iopjournal}

\usepackage{ragged2e}   
\usepackage{microtype}  

\usepackage{lineno}
\usepackage{amsmath,amssymb}
\usepackage{acro}
\usepackage{booktabs}    
\usepackage{siunitx}
\usepackage{adjustbox}

\DeclareAcronym{LVK}{
  short = LVK ,
  long = LIGO/Virgo/KAGRA ,
  short-plural =  ,
}
\DeclareAcronym{DESI}{
  short = DESI ,
  long = Dark Energy Spectroscopic Instrument ,
  short-plural =  ,
}
\DeclareAcronym{DR2}{
  short = DR2 ,
  long = Data Release 2 ,
  short-plural =  ,
}
\DeclareAcronym{BAO}{
  short = BAO ,
  long = baryon acoustic oscillations ,
  short-plural =  ,
}
\DeclareAcronym{BOSS}{
  short = BOSS ,
  long = Baryon Acoustic Oscillation Survey ,
  short-plural =  ,
}
\DeclareAcronym{6dFGS}{
  short = 6dFGS ,
  long = 6dF Galaxy Survey ,
  short-plural =  ,
}
\DeclareAcronym{SDSS-MGS}{
  short = SDSS-MGS ,
  long = Sloan Digital Sky Survey ``main galaxy sample'' ,
  short-plural =  ,
}
\DeclareAcronym{CMB}{
  short = CMB ,
  long = cosmic microwave background ,
  short-plural =  ,
}
\DeclareAcronym{ACT}{
  short = ACT ,
  long = Atacama Cosmology Telescope ,
  short-plural =  ,
}
\DeclareAcronym{SPT}{
  short = SPT-3G ,
  long = South Pole Telescope ,
  short-plural =  ,
}
\DeclareAcronym{S4}{
  short = S4 ,
  long = CMB Stage-IV ,
  short-plural =  ,
}
\DeclareAcronym{PTA}{
  short = PTA ,
  long = pulsar timing array ,
  short-plural = s ,
}
\DeclareAcronym{CSST}{
  short = CSST ,
  long = China Space Station Telescope ,
  short-plural =  ,
}

\begin{document}
\justifying             

\articletype{Paper} 

\title{New constraints on cosmological gravitational waves from CMB and BAO in light of dynamical dark energy}

\author{Sai Wang$^1$, Zhi-Chao Zhao$^{2^*}$}

\affil{$^1$School of Physics, Hangzhou Normal University, No.2318 Yuhangtang Road, Yuhang District, Hangzhou 311121, China}

\affil{$^{2^*}$Department of Applied Physics, College of Science, China Agricultural University, 17 Qinghua East Road, Haidian District, Beijing 100083, China.}

\email{zhaozc@cau.edu.cn, corrosponding author}

\keywords{cosmological gravitational waves, initial conditions, dynamical dark energy}

\begin{abstract}
\justifying             

In this work, we derive upper limits on the physical energy-density fraction today of cosmological gravitational waves, denoted by $\Omega_{\rm{gw}}h^{2}$, via analyzing \emph{Planck} \& ACT \& SPT CMB and DESI BAO data combination. In the standard cosmological model, we establish 95\% CL upper limits of  $\Omega_{\rm{gw}}h^{2} < 1.0 \times 10^{-6}$ for adiabatic initial conditions and $\Omega_{\rm{gw}}h^{2} < 2.7 \times 10^{-7}$ for homogeneous initial conditions, assuming a uniform prior for $\Omega_{\rm gw}h^{2}$. In light of dynamical dark energy, we get $\Omega_{\rm{gw}}h^{2} < 7.2 \times 10^{-7}$ (adiabatic) and $\Omega_{\rm{gw}}h^{2} < 2.4 \times 10^{-7}$ (homogeneous). In contrast, if a log-uniform prior was assumed for $\Omega_{\rm gw}h^{2}$, these constraints can become tighter by a factor of $\sim4$, suggesting the results to be prior-sensitive. Furthermore, we project the sensitivity achievable with LiteBIRD \& CMB Stage-IV measurements of CMB and CSST observations of BAO, forecasting 68\% CL uncertainties of $\sigma = 2.5 \times 10^{-7}$ (adiabatic) and $\sigma = 1.0 \times 10^{-7}$ (homogeneous) for ${\Omega_{\rm{gw}}h^{2}}$. The constraints we obtained in this work provide critical benchmarks for exploring the cosmological origins of gravitational waves within the frequency band $f \gtrsim 10^{-15}$\,Hz and potentially enable joint analysis with direct gravitational-wave detection sensitive to this regime. 

\end{abstract}

\section{Introduction}\label{sec:introduce}
Investigations of the cosmological gravitational-wave background bear profound significance. This background radiation predominantly originates from primordial gravitational-wave sources within the early universe, such as inflationary quantum fluctuations, coupled cosmological perturbations, first-order phase transitions, and topological defects, thus encoding critical information pertaining to the primordial universe (see, e.g., Ref.~\cite{Bian:2025ifp} for a review and references therein). Moreover, gravitational waves constitute linear cosmological tensor perturbations, whose evolutionary trajectory is governed by initial conditions, namely whether they manifest adiabatic or homogeneous characteristics \cite{Ma:1995ey,Bucher:1999re}.
{\color{black}  The adiabatic initial conditions apply when gravitational waves originate as a thermalized component from inflaton decay. In contrast, the homogeneous ones are the physically motivated choice for the unperturbed backgrounds generated by the most common sources, such as those mentioned above.} 
Therefore, upon detection of such gravitational waves, we shall not only decipher physical processes operative during the universe’s infancy, but also elucidate the fundamental mechanisms governing cosmic inception.

The cosmological gravitational-wave background constitutes a prime observational target for direct detection by gravitational-wave observatories. In the high-frequency regime, terrestrial interferometers have recently imposed rigorous upper bounds on the energy-density fraction spectrum of the stochastic gravitational-wave background \cite{KAGRA:2021kbb}. For the nanohertz-frequency band, multiple \ac{PTA} collaborations have reported compelling evidence for a stochastic background \cite{Xu:2023wog,EPTA:2023fyk,NANOGrav:2023gor,Reardon:2023gzh}. Regarding the millihertz-frequency band, space-borne gravitational-wave detectors are projected to conduct precision measurements of the stochastic background within the upcoming decade \cite{Hu:2017mde,Colpi:2024xhw,TianQin:2015yph}.

Complementary to direct detection of gravitational waves, cosmological probes employing the \ac{CMB} and \ac{BAO} measurements provide indirect constraints on the cosmological gravitational-wave background \cite{Smith:2006nka}. The \ac{CMB} detects such gravitational waves because they enhance the expansion rate of the universe during the epoch of photon decoupling. \ac{BAO} detects them because they suppress the growth of matter density perturbations. Leveraging the \emph{Planck} 2018 \ac{CMB} data alongside several \ac{BAO} datasets, Ref.~\cite{Clarke:2020bil} established observational upper bounds on the gravitational-wave energy density, depending on initial conditions which are either adiabatic or homogeneous. Were these constraints further integrated with \ac{PTA} datasets, they would yield tightened observational bounds on the gravitational waves from inflationary fluctuations \cite{Cabass:2015jwe,Liu:2015psa,Vagnozzi:2023lwo}, first-order phase transitions \cite{Bringmann:2023opz} and the induced gravitational waves \cite{Wang:2023sij,Zhu:2023gmx,Zhou:2024yke,Zhou:2025djn,Wu:2024qdb}.

The recently unveiled \ac{BAO} measurements from \ac{DESI} \ac{DR2} \cite{DESI:2025zgx} may prompt revisions to existing constraints on the cosmological gravitational-wave background. Notably, \ac{DESI} has not only delivered state-of-the-art \ac{BAO} measurements but also uncovered strong evidence for dynamical dark energy. Through synergistic analysis of these data with \emph{Planck} 2018 \ac{CMB} observations, Ref.~\cite{Wang:2025ljj} recently established an upper bound on the tensor-to-scalar ratio, which quantifies the spectral amplitude of primordial gravitational waves in the ultra-low frequency band, i.e., $f\lesssim10^{-16}$\,Hz. Nevertheless, in higher frequency bands, detailed investigations into the gravitational-wave energy density remain conspicuously absent in current literature, a gap designated as a paramount focus of the present study.

Next-generation \ac{CMB} and \ac{BAO} surveys hold considerable promise for probing the cosmological gravitational-wave background with unprecedented precision, thereby establishing significantly tighter constraints on the gravitational-wave energy density. Relative to extant facilities such as \emph{Planck} satellite, forthcoming observatories, notably the LiteBIRD satellite \cite{LiteBIRD:2022cnt} and \ac{S4} ground array \cite{CMB-S4:2016ple}, will deliver substantially refined measurements of the \ac{CMB} polarization. Similarly, next-generation spectroscopic surveys including the \ac{CSST} \cite{Gong:2019yxt} are anticipated to execute \ac{BAO} observations of enhanced spatial coverage and precision beyond \ac{DESI}'s capabilities. Through synergistic analysis of these datasets, we project stringent refinements to constraints on cosmological gravitational waves. This constitutes a complementary core objective of the present study.

This investigation pioneers a dual-pronged analytical framework, simultaneously constraining the gravitational-wave physical energy-density fraction today, denoted by $\Omega_{\mathrm{gw}}h^{2}$, through state-of-the-art cosmological observations and projecting detection thresholds for next-generation observatories. Incorporating dynamical dark energy's phenomenological consequences, we derive rigorous upper bounds on $\Omega_{\mathrm{gw}}h^{2}$ via combining the latest \ac{CMB} and \ac{BAO} data, under adiabatic versus homogeneous initial condition paradigms. These observationally derived constraints are subsequently benchmarked against the $\Lambda$CDM cosmology. We further quantify prospective sensitivity enhancements achievable through synergistic exploitation of next-generation \ac{CMB} measurements alongside \ac{CSST} \ac{BAO} observations. The resultant framework establishes novel upper bounds on cosmological gravitational-wave backgrounds across the $f \gtrsim 10^{-15}$ Hz domain, delineating definitive constraints for physical processes occurring in the early universe.

The paper is structured as follows. Section~\ref{sec:theory} presents the influence of cosmological gravitational waves on \ac{CMB} and \ac{BAO} under adiabatic and homogeneous initial conditions. Section~\ref{sec:presentconstraints} describes data analysis methods and derives upper limits on $\Omega_{\rm {gw}}h^{2}$ from current \ac{CMB} and \ac{BAO} observations. Section~\ref{sec:prospectivesensitivity} outlines cosmological forecasting methods to assess the sensitivity achievable with future \ac{CMB} and \ac{BAO} observations. Finally, Section~\ref{sec:summary} provides conclusions and discussion.

\section{Theory}\label{sec:theory}

Cosmological gravitational waves with frequencies above $10^{-15}$\,Hz can leave characteristic imprints on both the \ac{CMB} and matter power spectra, making them potentially constrained by \ac{CMB} and \ac{BAO} observations. On one hand, as a form of radiation, these gravitational waves enhance the cosmic expansion rate at photon decoupling, thereby altering the small-scale angular power spectrum of the \ac{CMB}. On the other hand, being free-streaming radiation, they suppress the growth of density perturbations, consequently modifying the matter power spectrum observable through \ac{BAO} measurements.

Gravitational waves, as cosmological tensor perturbations, satisfy evolution equations identical in form to those for massless neutrinos, but the solutions to these equations depend critically on the choice of initial conditions. 
{\color{black}  The necessary formalism involves deriving the linear perturbations of the Einstein and fluid-conservation equations (see details in, e.g., Ref~\cite{Clarke:2020bil}). The cosmic fluid includes not only photons, neutrinos, baryons, and dark matter, but also gravitational waves. The linear perturbations related to gravitational waves are characterized by the density ($\delta_{\rm gw}$), velocity ($\theta_{\rm gw}$), and shear ($\sigma_{\rm gw}$) perturbations in the synchronous gauge. Consequently, one obtains a system of four perturbed Einstein equations, supplemented by a corresponding set of fluid conservation equations for each individual species.
For the gravitational waves, which can be treated as a collisionless relativistic gas of gravitons, the fluid conservation equations are specifically given by \cite{Smith:2006nka,Clarke:2020bil}  
\begin{eqnarray}
    \dot{\delta}_{\rm gw} + \frac{4}{3}\theta_{\rm gw} + \frac{2}{3}\dot{h} &=& 0\, ,\label{eq:1}\\
    \dot{\theta}_{\rm gw} -\frac{1}{4}k^{2}\left(\delta_{\rm gw}-4\sigma_{\rm gw}\right) &=& 0\, ,\label{eq:2}\\
    \dot{\sigma}_{\rm gw} - \frac{2}{15}\left(2\theta_{\rm gw}+\dot{h}+6\dot{\eta}\right) &=& 0\, ,\label{eq:3}
\end{eqnarray}
where $h$ and $\eta$ standing for the scalar metric perturbations in the corresponding gauge, $k$ is the wavenumber of perturbations, and the overdot denotes the derivative with respect to the conformal time $\tau$. 
They are identical in form to those for massless neutrinos \cite{Smith:2006nka}. This equivalence underpins the use of cosmological observational data to constrain $\Omega_{\rm gw}$, with the phenomenology determined by the choice of initial conditions \cite{Smith:2006nka,Clarke:2020bil}.}

Initial conditions should be considered when we resolve the above equations. 
For linear cosmological perturbations, Ref.~\cite{Ma:1995ey} investigated adiabatic initial conditions, while Ref.~\cite{Bucher:1999re} explored non-adiabatic initial conditions, as explicitly demonstrated in Table~\ref{tab:0}. 
Under the former assumption, all matter components share identical fractional energy-density perturbations, allowing tensor perturbations to be treated equivalently to massless neutrino perturbations. This implies that the effect of gravitational waves is equivalent to that of the effective neutrino number, denoted by $N_{\rm eff}$. In contrast, under the latter non-adiabatic hypothesis, the cosmological gravitational-wave background exhibits no linear perturbations initially. Consequently, within the conformal Newtonian gauge, its energy-density perturbation must vanish. This is a distinct departure from the initial conditions for massless neutrinos, implying that the effect of gravitational waves can not be absorbed into $N_{\rm eff}$.

\begin{table*}[h]
\caption{Adiabatic and homogeneous initial conditions in the synchronous gauge, expanded to second order in $k\tau$, for the modes related to gravitational waves \cite{Clarke:2020bil}. We define $R_{i}=\rho_{i}/\sum\rho_{i}$ with $i$ representing $\gamma$, $\nu$, ${\rm gw}$. In this work, $\delta_{i}$, $\theta_{i}$, and $\sigma_{i}$, respectively, represent the density, velocity, and shear perturbations in this gauge with $h$ and $\eta$ denoting the metric perturbations. }\label{tab:0}
\begin{center}
\vspace{-2mm}\footnotesize 
\doublerulesep 0.2pt \tabcolsep 35pt
{\renewcommand{\arraystretch}{1.3}
\begin{tabular*}{\textwidth}{c|c|c}
\hline & Adiabatic I.C. & Homogeneous I.C. \\
\hline $h$ & $\frac{1}{2} k^2 \tau^2$ & $\frac{1}{2} k^2 \tau^2$ \\
$\eta$ & $1-\frac{\left(9-4 R_\gamma\right)}{12\left(19-4 R_\gamma\right)} k^2 \tau^2$ & $1-\frac{\left(9-4 R_\gamma+4 R_{\mathrm{gw}}\right)}{12\left(19-4 R_\gamma+4 R_{\mathrm{gw}}\right)} k^2 \tau^2$ \\
$\delta_\gamma$ & $-\frac{1}{3} k^2 \tau^2$ & $-\frac{R_{\mathrm{gw}}}{R_\gamma} \frac{20}{\left(19-4 R_\gamma+4 R_{\mathrm{gw}}\right)}$ \\
$\theta_\gamma$ & $\mathcal{O}\left(k^4 \tau^3\right)$ & $-\frac{R_{\mathrm{gw}}}{R_\gamma} \frac{5}{19-4 R_\gamma+4 R_{\mathrm{gw}}} k^2 \tau$ \\
$\delta_\nu$ & $-\frac{1}{3} k^2 \tau^2$ & $-\frac{1}{3} k^2 \tau^2$ \\
$\theta_\nu$ & $\mathcal{O}\left(k^4 \tau^3\right)$ & $\mathcal{O}\left(k^4 \tau^3\right)$ \\
$\sigma_\nu$ & $\frac{2}{3\left(19-4 R_\gamma\right)} k^2 \tau^2$ & $\frac{2}{3\left(19-4 R_\gamma+4 R_{\mathrm{gw}}\right)} k^2 \tau^2$ \\
$\delta_{\mathrm{gw}}$ & $-\frac{1}{3} k^2 \tau^2$ & $\frac{20}{19-4 R_\gamma+4 R_{\mathrm{gw}}}$ \\
$\theta_{\mathrm{gw}}$ & $\mathcal{O}\left(k^4 \tau^3\right)$ & $\frac{5}{19-4 R_\gamma+4 R_{\mathrm{gw}}} k^2 \tau$ \\
$\sigma_{\mathrm{gw}}$ & $\frac{2}{3\left(19-4 R_\gamma\right)} k^2 \tau^2$ & $\frac{4}{3\left(19-4 R_\gamma+4 R_{\mathrm{gw}}\right)} k^2 \tau^2$ \\
$\delta_{\mathrm{c}}$ & $-\frac{1}{4} k^2 \tau$ & $-\frac{1}{4} k^2 \tau$ \\
$\delta_{\mathrm{b}}$ & $-\frac{1}{4} k^2 \tau$ & $-\frac{R_\gamma\left(19-4 R_\gamma\right)+2 R_{\mathrm{gw}}\left(2 R_\gamma-5\right)}{4 R_\gamma(19-4 R_\gamma+4 R_{\mathrm{gw}})} k^2 \tau$ \\
\hline
\bottomrule[0.65pt] 
\end{tabular*}}
\end{center}
\end{table*}

\begin{figure}[h]
    \centering
    \includegraphics[width=1\linewidth]{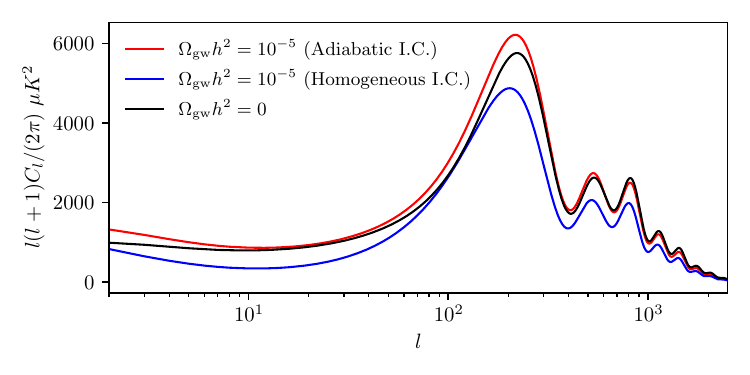}
    \caption{Effect of adiabatic versus homogeneous initial conditions on the CMB temperature angular power spectrum. Here, we consider only the $\Lambda$CDM model for the sake of illustration.  }
    \label{fig:0}
\end{figure}

The initial conditions of tensor perturbations affect the observational constraints on the cosmological gravitational waves from \ac{CMB} and \ac{BAO} measurements, as demonstrated by Figure~\ref{fig:0}. 
{\color{black}  Gravitational waves in the short-wave approximation can contribute an additional radiative energy component, increasing the expansion rate of the early universe and postponing the epoch of matter-radiation equality. This, in turn, shifts the sound horizon at the time of recombination, thereby modifying the power of CMB perturbations on small scales. On the other hand, as shown in Eqs.~(\ref{eq:1}, \ref{eq:3}), gravitational-wave perturbations can influence the time derivatives of the scalar metric perturbations $h$ and $\eta$. The latter, through the Einstein-Boltzmann equations, affects the evolution of photon and matter perturbations, thus leaving detectable imprints on the \ac{CMB} angular power spectrum.}
If adiabatic initial conditions are assumed, the impact of these gravitational waves on \ac{CMB} and \ac{BAO} is identical to that of massless neutrinos. Consequently, existing \ac{CMB}+\ac{BAO} constraints on $N_{\rm eff}$, can be directly translated into constraints on the energy density of gravitational waves. In contrast, under non-adiabatic initial conditions, the gravitational-wave imprint diverges from that of massless neutrinos. By analyzing \emph{Planck} 2018 \ac{CMB} data and several \ac{BAO} datasets, Ref.~\cite{Clarke:2020bil} derived upper limits (95\% confidence level) on the energy-density fraction of gravitational waves, i.e., $\Omega_{\mathrm{gw}}h^{2}<1.7\times10^{-6}$ for adiabatic initial conditions and $\Omega_{\mathrm{gw}}h^{2}<2.9\times10^{-7}$ for homogeneous initial conditions.

In this work, we derive constraints on the cosmological gravitational-wave background from the cutting-edge \ac{CMB} and \ac{BAO} observational data, and evaluate the detection capabilities of next-generation \ac{CMB} and \ac{BAO} facilities. Accordingly, we have implemented the evolution equations and initial conditions for gravitational waves in the \texttt{CLASS} \cite{Blas:2011rf} code. Specifically, within \texttt{CLASS}, we have replicated and adapted the code modules pertaining to massless neutrinos, adopting the initial conditions listed in Table~1 of Ref.~\cite{Clarke:2020bil} to enable the code for studying gravitational waves. We exclusively consider adiabatic and homogeneous initial conditions for gravitational waves, since isocurvature initial conditions are already tightly constrained by \emph{Planck} \cite{Planck:2018vyg}.

Having accounted for adiabatic and homogeneous initial conditions, we investigate the physical energy-density fraction of gravitational waves in both the standard cosmological model, i.e., the $\Lambda$CDM model, and its extension incorporating the Chevallier-Polarski-Linder (CPL) parametrization \cite{Chevallier:2000qy,Linder:2002et} of the equation of state of dark energy. For the former case, we consider seven independent parameters, namely $10^{6}\Omega_{\rm gw}h^{2}$, $\Omega_{b}h^{2}$, $\Omega_{c}h^{2}$, $100\theta_{\rm MC}$, $\tau$, $\ln(10^{10}A_{s})$, $n_{s}$, with the pivot scale being $k_{p}=0.05$\,${\rm Mpc}^{-1}$. The physical definitions of these parameters align with those established in the \emph{Planck} collaboration's paper \cite{Planck:2018vyg}. For the latter scenario, the inclusion of the dark-energy equation of state $w(a) = w_{0} + w_{a}(1-a)$, with $a$ being the scale factor of the universe, necessitates two additional independent parameters, specifically $w_{0}$ and $w_{a}$.


\section{Results from present datasets}\label{sec:presentconstraints}

To constrain the model parameters, we jointly analyze the cutting-edge \ac{CMB} and \ac{BAO} observational datasets. For the former, we employ the CMB-SPA data combination, whose composition has been detailed in Table~III of the SPT-3G collaboration's paper \cite{SPT-3G:2025bzu}. In brief, this refers to a data combination incorporating observations of \ac{CMB} temperature anisotropies, polarization, and lensing from \emph{Planck}, \ac{SPT}, and \ac{ACT} \cite{Carron:2022eyg,SPT-3G:2024atg,SPT-3G:2025bzu,SPT-3G:2025zuh,ACT:2025fju}. For the latter, we utilize the recent \ac{BAO} data from \ac{DESI} \cite{DESI:2025zgx}. Here, we perform cosmological parameter inference using the Markov-Chain Monte-Carlo (MCMC) sampler in \texttt{cobaya} \cite{Torrado:2020dgo}.

\begin{figure}[h]
    \centering
    \includegraphics[width=0.9\textwidth]{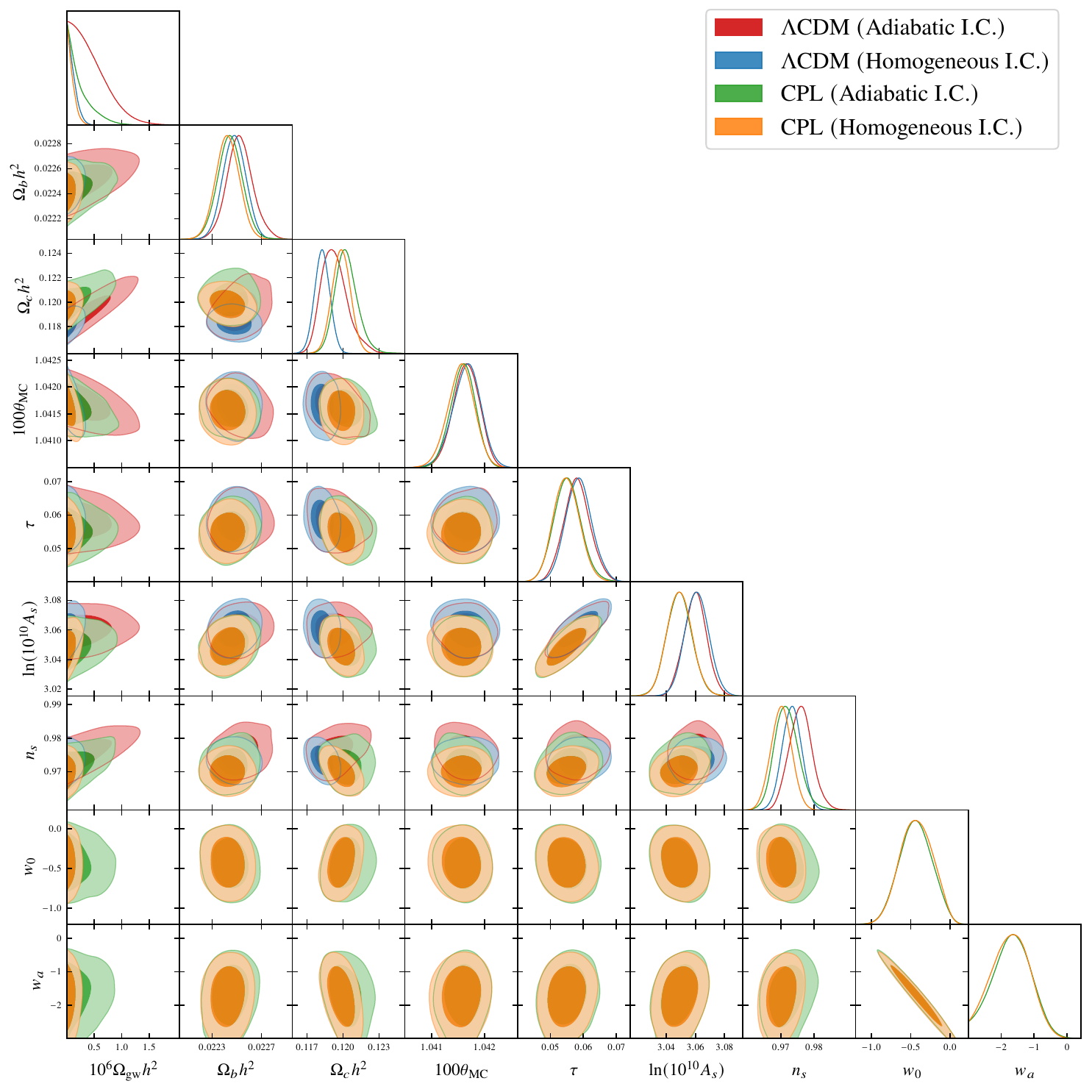}
    \caption{Same as Table~\ref{tab:1}, but we depict one- and two-dimensional posterior distributions of parameters. The dark and light shaded regions denote 68\% and 95\% confidence intervals, respectively. }
    \label{fig:1}
\end{figure}

The data analysis results are presented in Table~\ref{tab:1} and Figure~\ref{fig:1} for a uniform prior for $\Omega_{\rm gw}h^{2}$ in the interval of $0<\Omega_{\rm gw}h^{2}<4\times10^{-6}$ while in Table~\ref{tab:1-logUniform} and Figure~\ref{fig:1-logUniform} for a log-uniform prior for $\Omega_{\rm gw}h^{2}$ in the interval of $10^{-12}<\Omega_{\rm gw}h^{2}<4\times10^{-6}$. In this work, the abbreviation I.C. denotes initial conditions for gravitational waves. Parameter uncertainties are quoted at the 68\% confidence level in Tables~\ref{tab:1} and \ref{tab:1-logUniform}, while upper limits on parameters are reported at the 95\% confidence level. The one- and two-dimensional posterior distributions of these parameters are presented in Figures~\ref{fig:1} and \ref{fig:1-logUniform}.

\begin{table*}[h]
\caption{Parameter uncertainties at 68\% confidence level and upper limits at 95\% confidence level from analysis of CMB-SPA \cite{Carron:2022eyg,SPT-3G:2024atg,SPT-3G:2025bzu,SPT-3G:2025zuh,ACT:2025fju} and DESI BAO \cite{DESI:2025zgx}. Here, we use the uniform prior for $\Omega_{\rm gw}h^{2}$ in the interval of $0<\Omega_{\rm gw}h^{2}<4\times10^{-6}$. }\label{tab:1} 
\begin{center}
\vspace{-2mm}\footnotesize 
{\renewcommand{\arraystretch}{1.3}
\begin{tabular*}{\textwidth}{l|l|l|l|l}
      \hline
      & $\Lambda$CDM & $\Lambda$CDM & CPL & CPL\\
      & (Adiabatic I.C.) & (Homogeneous I.C.) & (Adiabatic I.C.) &
      (Homogeneous I.C.)\\
      \hline
      \textbf{$\Omega_{\mathrm{b}} h^2$} & $0.02253^{+ 0.00010}_{- 0.00011}$
      & $0.02248 \pm 0.00009$ & $0.02245 \pm 0.00010$ & $0.02242 \pm
      0.00010$\\
      \textbf{$\Omega_{\mathrm{c}} h^2$} & $0.11930^{+ 0.00085}_{- 0.00120}$ &
      $0.11827 \pm 0.00063$ & $0.12028^{+ 0.00080}_{- 0.00100}$ & $0.11986 \pm
      0.00076$\\
      \textbf{$100 \theta_{\textrm{\textrm{MC}}}$} & $1.04165 \pm 0.00025$ &
      $1.04166^{+ 0.00028}_{- 0.00024}$ & $1.04159 \pm 0.00023$ & $1.04156 \pm
      0.00025$\\
      \textbf{$\tau$} & $0.0581 \pm 0.0040$ & $0.0586 \pm 0.0040$ & $0.0552
      \pm 0.0041$ & $0.0549 \pm 0.0039$\\
      \textbf{$\ln (10^{10} A_{\mathrm{s}})$} & $3.0599 \pm 0.0077$ & $3.0610
      \pm 0.0081$ & $3.0490 \pm 0.0086$ & $3.0491 \pm 0.0085$\\
      \textbf{$n_{\mathrm{s}}$} & $0.9760 \pm 0.0035$ & $0.9733 \pm 0.0029$ &
      $0.9718^{+ 0.0032}_{- 0.0039}$ & $0.9702 \pm 0.0031$\\
      \textbf{$w_0$} & $-1$ (fixed) & $-1$ (fixed) & $- 0.44 \pm 0.20$ & $- 0.43
      \pm 0.20$\\
      \textbf{$w_a$} & $0$ (fixed) & $0$ (fixed) & $- 1.67 \pm 0.57$ & $- 1.71 \pm
      0.57$\\
      \textbf{$10^6 \Omega_{\mathrm{gw}} h^2$} & $< 1.04$ & $< 0.27$ & $<
      0.72$ & $< 0.24$\\
      \hline
\bottomrule[0.65pt] 
\end{tabular*}}
\end{center}
\end{table*}

{\color{black} 
\begin{table*}[h]
\caption{Same as Table \ref{tab:1}, but we use the log-uniform prior for $\Omega_{\rm gw}h^{2}$ in the interval of $10^{-12}<\Omega_{\rm gw}h^{2}<4\times10^{-6}$. }\label{tab:1-logUniform} 
\begin{center}
\vspace{-2mm}\footnotesize 
{\renewcommand{\arraystretch}{1.3}
\begin{tabular*}{\textwidth}{l|l|l|l|l}
      \hline
      & $\Lambda$CDM & $\Lambda$CDM & CPL & CPL\\
      & (Adiabatic I.C.) & (Homogeneous I.C.) & (Adiabatic I.C.) &
      (Homogeneous I.C.)\\
      \hline
  \textbf{$\Omega_{\mathrm{b}} h^2$} & $0.02247 \pm 0.00009$ & $0.02246 \pm
  0.00009$ & $0.02241 \pm 0.00010$ & $0.02241 \pm 0.00009$\\
  \textbf{$\Omega_{\mathrm{c}} h^2$} & $0.11816^{+ 0.00059}_{- 0.00070}$ &
  $0.11807 \pm 0.00060$ & $0.11969 \pm 0.00077$ & $0.11967 \pm 0.00075$\\
  \textbf{$100 \theta_{\text{\rm{MC}}}$} & $1.04179 \pm 0.00022$ &
  $1.04180 \pm 0.00023$ & $1.04166 \pm 0.00023$ & $1.04167 \pm 0.00023$\\
  \textbf{$\tau$} & $0.0586 \pm 0.0040$ & $0.0586 \pm 0.0040$ & $0.0549 \pm
  0.0040$ & $0.0550 \pm 0.0039$\\
  \textbf{$\ln (10^{10} A_{\mathrm{s}})$} & $3.0583 \pm 0.0079$ & $3.0581
  \pm 0.0079$ & $3.0469 \pm 0.0082$ & $3.0470 \pm 0.0082$\\
  \textbf{$n_{\mathrm{s}}$} & $0.9732 \pm 0.0030$ & $0.9730 \pm 0.0029$ &
  $0.9701 \pm 0.0030$ & $0.9699 \pm 0.0030$\\
  \textbf{$w_0$} & $- 1$ (fixed) & $- 1$ (fixed) & $- 0.42 \pm 0.20$ & $-
  0.42 \pm 0.20$\\
  \textbf{$w_a$} & $0$ (fixed) & $0$ (fixed) & $- 1.72 \pm 0.56$ & $- 1.73
  \pm 0.56$\\
  \textbf{$10^6 \Omega_{\mathrm{gw}} h^2$} & $< 0.28$ & $< 0.07$ &
  $< 0.13$ & $< 0.06$\\
  \hline
\bottomrule[0.65pt] 
\end{tabular*}}
\end{center}
\end{table*}
}

{\color{black} 
\begin{figure}[h]
    \centering
    \includegraphics[width=0.9\textwidth]{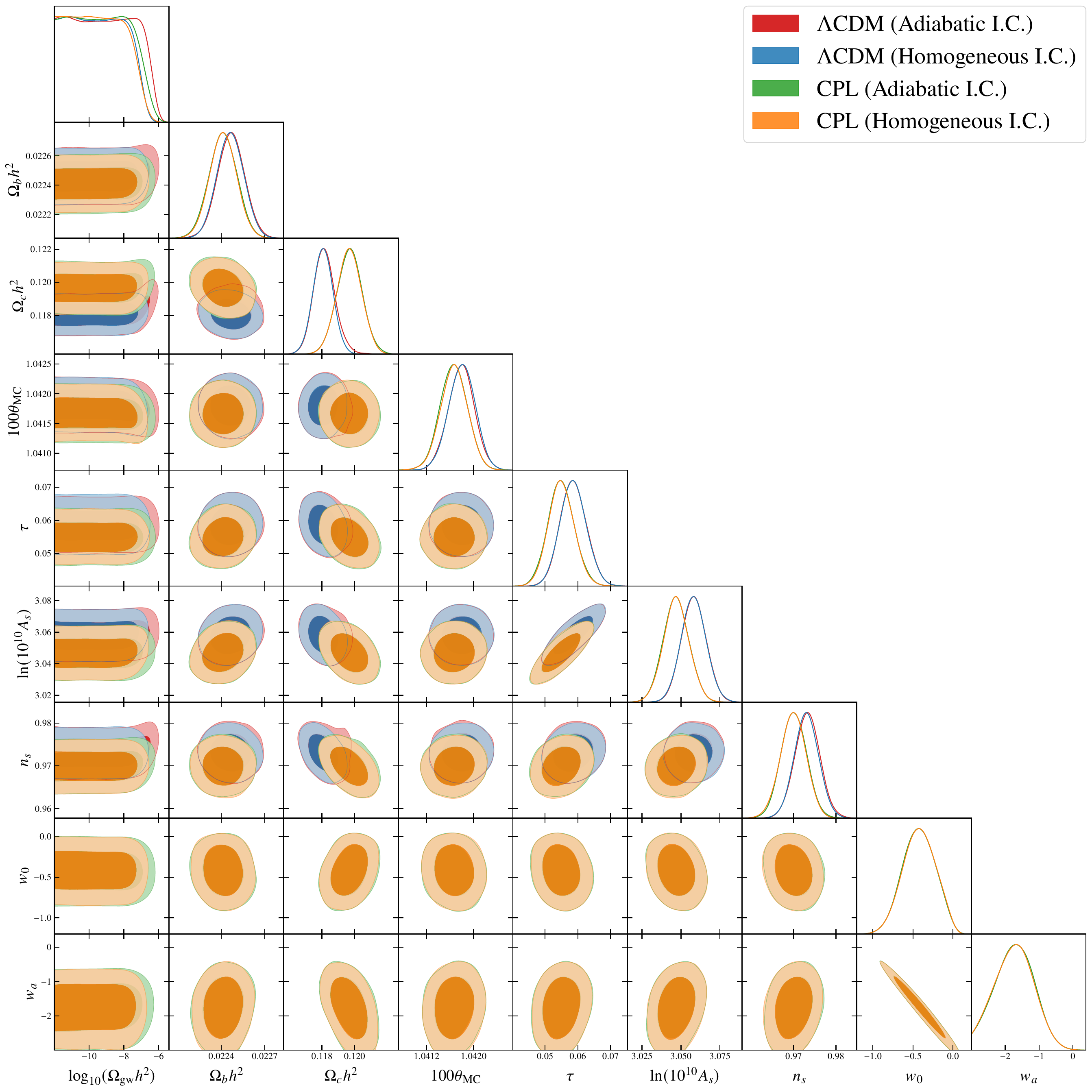}
    \caption{Same as Table~\ref{tab:1-logUniform}, but we depict one- and two-dimensional posterior distributions of parameters. The dark and light shaded regions denote 68\% and 95\% confidence intervals, respectively. }
    \label{fig:1-logUniform}
\end{figure}
}

Based on Table~\ref{tab:1}, we find that the addition of dynamical dark energy suppresses the observational upper bounds on $\Omega_{\rm gw}h^{2}$, with the degree of suppression depending on the chosen initial conditions for gravitational waves. Under adiabatic initial conditions, the upper limits are $\Omega_{\rm gw}h^{2}<1.0\times10^{-6}$ ($\Lambda$CDM) and $\Omega_{\rm gw}h^{2}<7.2\times10^{-7}$ (CPL), revealing a difference of $\sim$30\%. For homogeneous initial conditions, they become $\Omega_{\rm gw}h^{2}<2.7\times10^{-7}$ ($\Lambda$CDM) and $\Omega_{\rm gw}h^{2}<2.4\times10^{-7}$ (CPL), revealing a difference of $\sim$10\%. However, these results indicate that the constraints on $\Omega_{\rm gw}h^{2}$ under homogeneous initial conditions are less sensitive to the choice of dark energy model. This is also revealed by the one-dimensional posterior distributions of $\Omega_{\rm gw}h^{2}$, as shown in the top subfigure of Figure~\ref{fig:1}. Furthermore, we find that for the $\Lambda$CDM model, the upper limits obtained here are tighter by at most 40\% than those reported in existing literature \cite{Clarke:2020bil}. This improvement is primarily attributable to our employment of state-of-the-art observational datasets.

{\color{black}  A comparison between the results in Table~\ref{tab:1-logUniform} and Table~\ref{tab:1} reveals that our cosmological constraints on cosmological gravitational waves exhibit a non-negligible dependence on the choice of parameter prior. The bounds derived under a log-uniform prior are more stringent by a factor of $\sim4$ than those obtained using a uniform prior. To be specific, under adiabatic initial conditions, the upper limits are given by $\Omega_{\rm gw}h^{2}<2.8\times10^{-7}$ ($\Lambda$CDM) and $\Omega_{\rm gw}h^{2}<1.3\times10^{-7}$ (CPL), while for homogeneous initial conditions, they become $\Omega_{\rm gw}h^{2}<0.7\times10^{-7}$ ($\Lambda$CDM) and $\Omega_{\rm gw}h^{2}<0.6\times10^{-7}$ (CPL). Since $\Omega_{\rm gw}$ can span several orders of magnitude, a log-uniform prior provides a more natural and less informative sampling across this broad range. Consequently, it more effectively constrains the parameter's extreme values. This comparison indicates that while the numerical results are prior-sensitive, the stricter constraints derived under the log-uniform prior represent a more robust and physically credible conclusion of our analysis.}

We further find that homogeneous initial conditions yield more stringent observational constraints on $\Omega_{\rm gw}h^{2}$ compared to adiabatic initial conditions. This could be attributed to the fact that while adiabatic gravitational waves enhance the total power of cosmological perturbations, their homogeneous counterparts suppress it, as demonstrated in Figure~\ref{fig:0}. Physically, under homogeneous initial conditions, gravitational-wave perturbations and photon perturbations evolve out of phase within the horizon due to their initial conditions having opposite signs.

\section{Sensitivity of future facilities}\label{sec:prospectivesensitivity}

The aforementioned findings can be rigorously tested by future cosmological observational data and are expected to enable more precise measurements of the cosmological gravitational-wave background. As a next-generation flagship spectroscopic survey, \ac{CSST} \cite{Gong:2019yxt} is expected to obtain slitless spectroscopic data from numerous galaxies and AGNs, potentially delivering enhanced measurements of \ac{BAO} across multiple redshift bins, thereby providing deeper insights into the nature of dark energy. Furthermore, a joint analysis of \ac{CSST}'s \ac{BAO} measurements with the \ac{CMB} observational data from next-generation experiments such as LiteBIRD \cite{LiteBIRD:2022cnt} and \ac{S4} \cite{CMB-S4:2016ple} could achieve higher-precision constraints on cosmological parameters, particularly including $\Omega_{\rm gw}h^{2}$.

By utilizing the MCMC sampler in the \texttt{MontePython} \cite{Brinckmann:2018cvx,Audren:2012wb} code, we analyze the combination of \ac{BAO} data from \ac{CSST} and \ac{CMB} data from LiteBIRD and \ac{S4} to determine the projected precision of $\Omega_{\rm gw}h^{2}$ measurements from this experimental configuration. For the former, we adopt the pessimistic precision of \ac{CSST} \ac{BAO} measurements from Table~3 in Ref.~\cite{Miao:2023umi}, representing a conservative projection. We integrate this dataset with its likelihood into \texttt{MontePython} via calibrating the dataset to align with our fiducial models. For the latter, we follow the methodology of Ref.~\cite{Brinckmann:2018cvx}, i.e., using LiteBIRD's projected precision for data at large angular scales while adopting \ac{S4}'s projected precision for data at small angular scales. Specifically, we employ the mock likelihoods \texttt{litebird\_lowl}, integrating measurements of temperature anisotropies and polarization at $2\leq\ell\leq50$, and \texttt{cmb\_s4\_highl}, integrating measurements of temperature anisotropies, polarization, and lensing at $\ell>50$. As an approximation, we neglect potential correlations between \ac{CMB} and \ac{BAO} measurements. Furthermore, each fiducial model is determined by the upper-limit value of $\Omega_{\rm gw}h^{2}$ and the central values of other parameters provided in Table~\ref{tab:1}. 
Here, we still assume a uniform prior for this parameter in order to perform a conservative estimate.

\begin{figure}[h]
    \centering
    \includegraphics[width=0.7\linewidth]{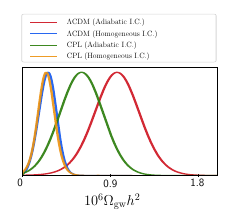}
    \caption{One-dimensional posterior distributions of $\Omega_{\rm gw}h^{2}$ obtained from the data combination of CSST \cite{Gong:2019yxt}, LiteBIRD \cite{LiteBIRD:2022cnt}, and S4 \cite{CMB-S4:2016ple}. Each fiducial model is determined by the upper-limit value of $\Omega_{\rm gw}h^{2}$ and the central values of other parameters provided in Table~\ref{tab:1}. }
    \label{fig:2}
\end{figure}

The data analysis reveals that the combined dataset is projected to detect deviations of $\Omega_{\rm gw}h^{2}$ from zero at the $\sim2.5-4\sigma$ confidence level; otherwise, it is expected to further tighten the observational upper limits on $\Omega_{\rm gw}h^{2}$. Under adiabatic initial conditions, the projected results (68\% confidence interval) are $\Omega_{\rm gw}h^{2}=1.04_{-0.25}^{+0.24}\times10^{-6}$ ($\Lambda$CDM) and $\Omega_{\rm gw}h^{2}=7.2_{-2.5}^{+2.3}\times10^{-7}$ (CPL), with the measurement precisions being almost the same. For homogeneous initial conditions, they become $\Omega_{\rm gw}h^{2}=2.7_{-1.0}^{+0.9}\times10^{-7}$ ($\Lambda$CDM) and $\Omega_{\rm gw}h^{2}=2.4_{-1.0}^{+0.9}\times10^{-7}$ (CPL), with the measurement precisions also being the same. Correspondingly, the one-dimensional posterior distributions of this parameter are presented in Figure~\ref{fig:2}, which allows for direct comparison with the top subfigure of Figure~\ref{fig:1}.

It is particularly intriguing that across both the $\Lambda$CDM and CPL cosmological frameworks, the projected precision for measuring $\Omega_{\rm gw}h^{2}$ with future detection facilities registers at $\sigma\simeq2.5\times10^{-7}$ (and $2\sigma\simeq4.8\times10^{-7}$) under adiabatic initial conditions and $\sigma\simeq1.0\times10^{-7}$ (and $2\sigma\simeq1.8\times10^{-7}$) under homogeneous initial conditions, respectively, revealing model-independent enhancement prospects. Such precision enhancements stem from the deployment of more-sensitive next-generation instrumentation, surpassing current measurement capabilities. 
Furthermore, being consistent with analysis shown in the last section, homogeneous initial conditions still produce tighter observational constraints on $\Omega_{\rm gw}h^{2}$ than adiabatic initial conditions.

\section{Summary and discussion}\label{sec:summary}

In this study, we have derived new constraints on the present-day physical energy-density fraction of cosmological gravitational waves through joint analysis of the data combination of \ac{CMB} from \emph{Planck} \& \ac{ACT} \& \ac{SPT} and \ac{BAO} from \ac{DESI}.  As presented in Table~\ref{tab:1}, under physically-motivated homogeneous initial conditions for tensor perturbations, the constraints at the 95\% confidence level were measured as $\Omega_{\rm gw}h^{2}<2.7\times10^{-7}$ in the $\Lambda$CDM model and $\Omega_{\rm gw}h^{2}<2.4\times10^{-7}$ in the CPL model, respectively. However, under adiabatic initial conditions, these constraints became less stringent. {\color{black}  Furthermore, the choice of priors for $\Omega_{\rm gw}h^{2}$ can change the results of our present work to some extent. For the aforementioned upper limits, we have used the uniform prior for $\Omega_{\rm gw}h^{2}>0$. In contrast, if we used the log-uniform prior alternatively, the upper limits, as presented in Table~\ref{tab:1-logUniform}, would become smaller by a factor of $\sim4$, leading to more stringent contratints.}

{\color{black}  When jointly analyzing \ac{CMB} and \ac{BAO} data, we should notice potential residual systematics that may not have been fully eliminated, as well as possible correlations between \ac{CMB} and \ac{BAO} measurements. 
For the \ac{CMB} anisotropies and polarization, correlations between \emph{Planck} and \ac{ACT} are controlled via the multipole cuts, while \ac{SPT} is effectively independent of \emph{Planck} and \ac{ACT}, given minimal sky overlap \cite{ACT:2025fju}. For the \ac{CMB} lensing, correlations between \ac{ACT} and \ac{SPT} have been shown negligible for cosmological parameter estimation \cite{SPT-3G:2025bzu}. For the \ac{BAO} distances, DESI DR2 has propagated a quantified systematics budget in the covariance and demonstrated that strong correlations between redshift-bin systematics have no detectable impact on cosmological constraints \cite{DESI:2025zgx}. Furthermore, \ac{CMB} and \ac{BAO} are typically modeled to be independent in joint fits. Ref.~\cite{Kou:2025hvg} has also shown that cross-correlations between \ac{CMB} and \ac{BAO} induce only small corrections for most parameters, so neglecting such cross-correlations is a well-justified approximation. Therefore, we expect the results of our present work to remain robust.}

We have further derived projected constraints on $\Omega_{\rm gw}h^{2}$ by analyzing the combined \ac{CMB} data from LiteBIRD and \ac{S4} alongside \ac{BAO} data from \ac{CSST}. These next-generation facilities were anticipated to improve current observational bounds on $\Omega_{\rm gw}h^{2}$ due to their enhanced detection capabilities. We revealed that homogeneous initial conditions deliver enhanced measurement precision for $\Omega_{\rm gw}h^{2}$ compared to adiabatic initial conditions, while the specific nature of dark energy exerts no discernible influence on the precision. If we still used a uniform prior of $\Omega_{\rm gw}h^{2}$ as a conservative estimate, the measurement precision was shown as $\sigma\simeq2.5\times10^{-7}$ (and $2\sigma\simeq4.8\times10^{-7}$) under adiabatic initial conditions and $\sigma\simeq1.0\times10^{-7}$ (and $2\sigma\simeq1.8\times10^{-7}$) under homogeneous initial conditions, respectively.

In this work, we have aimed to establish model-agnostic constraints on the cosmological gravitational-wave background. Consequently, we did not focus on specific scenarios of cosmological gravitational waves. The constraints obtained here could provide critical benchmarks for exploring the cosmological origins of gravitational waves within the frequency band $f \gtrsim 10^{-15}$\,Hz and potentially enable joint analysis with direct gravitational-wave detection sensitive to this regime. For example, recent data from \acp{PTA} have provided strong evidence for a stochastic gravitational-wave background in the nanohertz frequency range \cite{Xu:2023wog,EPTA:2023fyk,NANOGrav:2023gor,Reardon:2023gzh}. If this signal originates from cosmological sources, such as induced gravitational waves, its infrared (IR) spectral behavior could offer a viable explanation (see, e.g., Ref.~\cite{NANOGrav:2023hvm}). The constraints derived in this work, while less direct than \ac{PTA} measurements in the nHz band, offer a complementary probe. \acp{PTA} measure the spectral energy density at nanohertz frequencies, whereas our cosmological analysis is sensitive to the total energy density integrated over a much broader frequency range. This complementarity means that, in principle, the combination of cosmological and \ac{PTA} data can break degeneracies and provide significantly tighter constraints on models of the cosmological gravitational-wave background \cite{Bringmann:2023opz,Wang:2023sij,Zhu:2023gmx}. Should the need have arisen, we can combine the cosmological constraints with direct gravitational-wave detection data for in-depth model investigations. However, we would like to designate such research for future works.

\section{acknowledgments}

S.W. is supported by the National Key R\&D Program of China No. 2023YFC2206403 and the National Natural Science Foundation of China (Grant No. 12175243).  
Z.C.Z. is supported by the National Key Research and Development Program of China Grant No. 2021YFC2203001.

\bibliographystyle{JHEP}
\bibliography{biblio}

\end{document}